\documentclass[nohyperref]{article}

\usepackage{microtype}
\usepackage{graphicx}
\usepackage{subfigure}
\usepackage{booktabs} 

\usepackage{hyperref}

\usepackage[accepted]{icml2022}

\usepackage{amsmath}
\usepackage{amssymb}
\usepackage{mathtools}
\usepackage{amsthm}

\usepackage{aas_macros}

\usepackage[capitalize,noabbrev]{cleveref}

\theoremstyle{plain}

\theoremstyle{definition}

\theoremstyle{remark}

\usepackage[textsize=tiny]{todonotes}

\icmltitlerunning{Neural Network Selection Correction in Strong Lensing}

\begin{document}

\twocolumn[
\icmltitle{Population-Level Inference of Strong Gravitational Lenses with Neural Network-Based Selection Correction}




\icmlsetsymbol{equal}{*}

\begin{icmlauthorlist}
\icmlauthor{Ronan Legin}{udem}
\icmlauthor{Connor Stone}{udem,queen,mila}
\icmlauthor{Yashar Hezaveh}{udem,cca}
\icmlauthor{Laurence Perreault-Levasseur}{udem,mila,cca}
\end{icmlauthorlist}

\icmlaffiliation{cca}{Center for Computational Astrophysics, Flatiron Institute, NY, USA}
\icmlaffiliation{udem}{Department of Physics, Universit\'{e} de Montr\'{e}al, Montr\'{e}al, Qu\'{e}bec, Canada}
\icmlaffiliation{queen}{Department of Physics, Engineering Physics and Astronomy, Queen’s University, Kingston, ON K7L 3N6, Canada}
\icmlaffiliation{mila}{Mila - Quebec Artificial Intelligence Institute, Montr\'{e}al, Qu\'{e}bec, Canada}

\icmlcorrespondingauthor{Ronan Legin}{ronan.legin@umontreal.ca}

\icmlkeywords{Machine Learning, ICML}

\vskip 0.3in
]



\printAffiliationsAndNotice{} 

\begin{abstract}

A new generation of sky surveys is poised to provide unprecedented volumes of data containing hundreds of thousands of new strong lensing systems in the coming years. Convolutional neural networks are currently the only state-of-the-art method that can handle the onslaught of data to discover and infer the parameters of individual systems. However, many important measurements that involve strong lensing require population-level inference of these systems. In this work, we propose a hierarchical inference framework that uses the inference of individual lensing systems in combination with the selection function to estimate population-level parameters. In particular, we show that it is possible to model the selection function of a CNN-based lens finder with a neural network classifier, enabling fast inference of population-level parameters without the need for expensive Monte Carlo simulations. 

\end{abstract}

\section{Introduction}

Strong gravitational lensing is a phenomenon where light rays of a distant background source are  deflected by the gravity of foreground matter, resulting in the production of multiple images. These lensing systems can be used for important applications in cosmology, such as constraining the mass of dark matter present in foreground galaxies \citep[e.g.,][]{2012Natur.481..341V, 2016ApJ...823...37H}, providing precise  measurements of the expansion rate of the universe \citep[e.g.,][]{2020MNRAS.498.1420W} and learning more about the morphology of distant high redshift galaxies \citep[e.g.,][]{2013ApJ...762...32C}.

In the next few years, a new generation of wide-area sky surveys from the Rubin Observatory's Legacy Survey of Space and Time (LSST) and the Euclid Space Telescope (EST) will provide an unprecedented volume of data containing hundreds of thousands of new strong lensing systems \citep{2015ApJ...811...20C}.

Recent works have focused on developing new and efficient computational methods to analyze these large sets of data. In particular, machine learning has been proposed to efficiently perform various strong lensing analysis tasks. Convolutional neural networks (CNN) have been trained in classification tasks to detect strong lenses from survey data \citep{2018A&A...611A...2S, 2018MNRAS.473.3895L}. They have been used to discover new strong lensing systems from the Kilo Degree Survey \citep[KiDS]{2020ApJ...899...30L} and the Ultraviolet Near Infrared Optical Northern Survey~\citep{Savary2021}, and have generally surpassed traditional methods in terms of speed and accuracy according to benchmark tests \citep{2019A&A...625A.119M}. These networks have also been used to perform fast and automated inference of the lensing parameters from the observations of individual strong lensing systems  \citep{2017Natur.548..555H,  2017ApJ...850L...7P,2021ApJ...909..187W, 2021ApJ...910...39P, 2021arXiv211205278L}, speeding up the analysis by more than $\sim10$ million times compared to traditional methods.

However, many important science applications involving strong gravitational lenses rely on population-level statistics of lensing systems \citep[e.g.,][]{2021A&A...651A..18S,  2021A&A...656A.153S, 2022A&A...659A.133S, 2022A&A...659A.132S}. Generally, the task of population-level inference is best framed within a hierarchical framework, where the problem is broken down in multiple stages. In a hierarchical framework, a clear separation between the different components of the analysis is provided, (e.g., between the theoretical models underlying the population distribution and the parameters describing the individual observations themselves). Generally, the problem includes the inference of global hyperparameters $\lambda$ describing the population distribution of model parameters $\theta$ i.e the parameters describing individual strong lenses. It is important, however, for these hierarchical models to take into account the selection mechanisms of the observed data sets to avoid biased inference results \citep{2021A&A...651A..18S, 2022A&A...659A.132S}. The selection function would depend on the model parameters of individual strong lensing observations. Traditionally, this makes population-level inference difficult in standard settings, where complex selections are difficult to quantify and the corrections require a large number of simulations at every step of the posterior sampling process. 

In this work, we model the selection function of a CNN-based strong lensing detector using neural networks. We propose an efficient and flexible likelihood-free inference framework for performing accurate large scale population-level inference of strong lensing parameters that can readily incorporate learned selection effects. In addition, our framework circumvents issues regarding the interpretability of typical black-box inference machines by allowing clear separation between the different components of our hierarchical population model. In section \ref{methods}, we describe our Bayesian hierarchical inference framework for inferring the posterior distribution of hyperparameters $\lambda$ describing the population of strong gravitational lensing model parameters $\theta$. We also detail our strong lensing simulations and the modeling of the CNN-based selection bias using neural networks. In section \ref{results}, we test our hierarchical method to infer the population of strong lensing model parameters based on a set of 1000 simulated strong lensing observations.

\section{Methods}
\label{methods}
Given a set of observations $\{x_i\}$, the task is to infer the posterior distribution of population hyperparameters $p(\lambda|\{x_i\})$. Using Bayes' theorem, this can be expressed as

\begin{align}
    p(\lambda|\{x_i\}) &=  \frac{p(\lambda) \prod_{i} p(x_i|\lambda)}{\int d\lambda^{\prime} p(\lambda^{\prime}) \prod_{i}
    p(x_i|\lambda^{\prime})} \nonumber \\
    &= \left[\int d\lambda^{\prime} \frac{p(\lambda^{\prime})}{p(\lambda)} \prod_{i} \frac{p(x_i|\lambda^{\prime})}{p(x_i|\lambda)}\right]^{-1} \label{pop_posterior},
\end{align}

In a typical hierarchical model, the individual observations $x_i$ may explicitly depend on intermediate model parameters $\theta$ and only implicitly on $\lambda$. In this case, the likelihood $p(x_i|\lambda)$ can be expanded as  

\begin{align}
    \label{hier}
    p(x_i|\lambda) &= \int p(x_i|\theta) p(\theta| \lambda) d\theta,
\end{align}

where $\lambda$ is ignored in $p(x_i|\theta)$ as it only implicitly affects the observation $x_i$.

In practice, it may be more likely to observe certain events $x_i$. This may lead to the chosen set of events $\{x_i\}$ to instead represent a biased selection of all possible observations, which would consequently result in a biased inference of the posterior distribution of the population hyperparameters $p(\lambda|\{x_i \})$. In the presence of selection effects \citep{2019MNRAS.486.1086M, 2021PhRvD.104h3008M}, the proper expression for the likelihood $p(x_i|\lambda)$ is

\begin{align}
    \label{biased_likelihood}
    p(x_i|\lambda) &=  \frac{\int p(x_i|\theta) p(\theta| \lambda) d\theta}{\alpha(\lambda)},
\end{align}

where $\alpha(\lambda)$ is evaluated by integrating over the product of the probability of detection $p_{\mathrm{det}}(\theta)$ of an event with model parameters $\theta$ and the population model $p(\theta|\lambda)$,

\begin{align}
    \label{alpha_eq}
    \alpha(\lambda) =  \int p_{\mathrm{det}}(\theta) p(\theta| \lambda) d\theta.
\end{align}

As such, accounting for selection bias incorporates an  overall normalization factor $\alpha(\lambda)$ within the likelihood $p(x|\lambda)$ that can be interpreted as the fraction of samples from the population model $p(\theta|\lambda)$ that would be observed given the selection function $p_{\mathrm{det}}(\theta)$ \citep{2019MNRAS.486.1086M}.  Note that Equation \ref{biased_likelihood} can be rewritten as

\begin{align}
    \label{biased_likelihood_exp}
    p(x_i|\lambda) &= \mathbb{E}_{p(\theta|x_i)} \left[ \frac{p(\theta|\lambda)}{p(\theta)} \right] \frac{p(x_i)}{\alpha(\lambda)},
\end{align}

by applying Bayes' theorem on the likelihood of the individual lensing observations $p(x_i|\theta)$ to obtain an expectation value over the lensing model parameter posterior $p(\theta|x_i)$.

Plugging this new expression back into Equation \ref{pop_posterior}, we obtain 

{\small
\begin{align}
    \label{pop_posterior_final}
    p(\lambda|\{x_i\}) &=  \left[\int d\lambda^{\prime} \frac{p(\lambda^{\prime})}{p(\lambda)} \prod_{i} \frac{\mathbb{E}_{p(\theta|x_i)} \left[ \frac{p(\theta|\lambda^{\prime})}{p(\theta)\alpha(\lambda^{\prime})} \right]}{\mathbb{E}_{p(\theta|x_i)} \left[ \frac{p(\theta|\lambda)}{p(\theta)\alpha(\lambda)} \right]}\right]^{-1},
\end{align}}

where the marginal probabilities $p(x_i)$ from applying Bayes' theorem on both $p(x_i|\lambda)$ and $p(x_i|\lambda^{\prime})$ cancel out.
As the expectation values cannot be evaluated analytically, we can instead approximate them as a mean over samples from the lensing posterior $p(\theta|x_i)$. Similarly, we can estimate $\alpha(\lambda)$ by rewriting Equation \ref{alpha_eq} as the expectation value of $p_{\mathrm{det}}(\theta)$ over $p(\theta|\lambda)$ and approximate it with samples from $p(\theta|\lambda)$.

The difficulty in estimating $\alpha(\lambda)$, however, is that the selection function $p_{\mathrm{det}}(\theta)$, which captures the effects of the selection bias, is typically complex and not explicitly known \citep[e.g.,][]{2022A&A...659A.132S}. This is especially true if the selection method is based on CNN detections of strong lenses, as the entire detection algorithm is framed as a black-box.

In this work, we circumvent this issue by modeling the selection function $p_{\mathrm{det}}(\theta)$ of CNN-based detectors of strong lenses. In particular, we train a neural network to classify between detected and non-detected samples of strong lensing model parameters $\theta$. The output probability of this classification network results in the probability of detection $p_{\mathrm{det}}(\theta)$ of a lensing system with parameters $\theta$.

Furthermore, we train a convolutional Bayesian Neural Network (BNN) to predict the posterior distribution of strong lensing model parameters $p(\theta|x)$ and to rapidly sample from it in order to approximate the expectation values from Equation \ref{pop_posterior_final}. Given that upcoming sky surveys are estimated to discover hundreds of thousands of new strong lenses, traditional methods such as likelihood-based inference, which require time-consuming analyses from strong lensing experts, will be intractable for this task. Instead, BNNs have been shown to be orders of magnitude faster than traditional methods and can provide well-calibrated uncertainties on strong lensing parameter predictions \citep{2017ApJ...850L...7P, 2021ApJ...909..187W, 2021ApJ...910...39P}.

\subsection{Simulations}

We simulate two sets of strong lensing observations: low-resolution images similar to what is expected to be observed by wide-area surveys such as LSST and the corresponding high-resolution follow-up observations. As a whole, our simulations can be described by a 13-dimensional model parameter vector $\theta$. Examples of the lensing simulations are shown in appendix \ref{appendix}. 

At test time, the CNN detector is applied to the set of mock LSST lensing observations. Then, for the images that are classified as a detection, we apply our BNN model to their follow-up observations in order to infer the posterior distribution of strong lensing model parameters $p(\theta|x)$ with higher precision. Given the inferred posterior $p(\theta|x)$, we can evaluate Equation \ref{pop_posterior_final} by approximating the expectation values using samples from it.

\subsection{Selection function}
\label{methods_selec}
We train a multilayer perceptron (MLP) neural network to predict the probability of detection $p_{\mathrm{det}}(\theta)$ for CNN-based detectors of strong gravitational lenses. This neural network is trained to classify between a set of $\theta$ that results in the detection of strong lensing systems and a set that results in a non-detection.

\begin{figure}
\includegraphics[width=8cm]{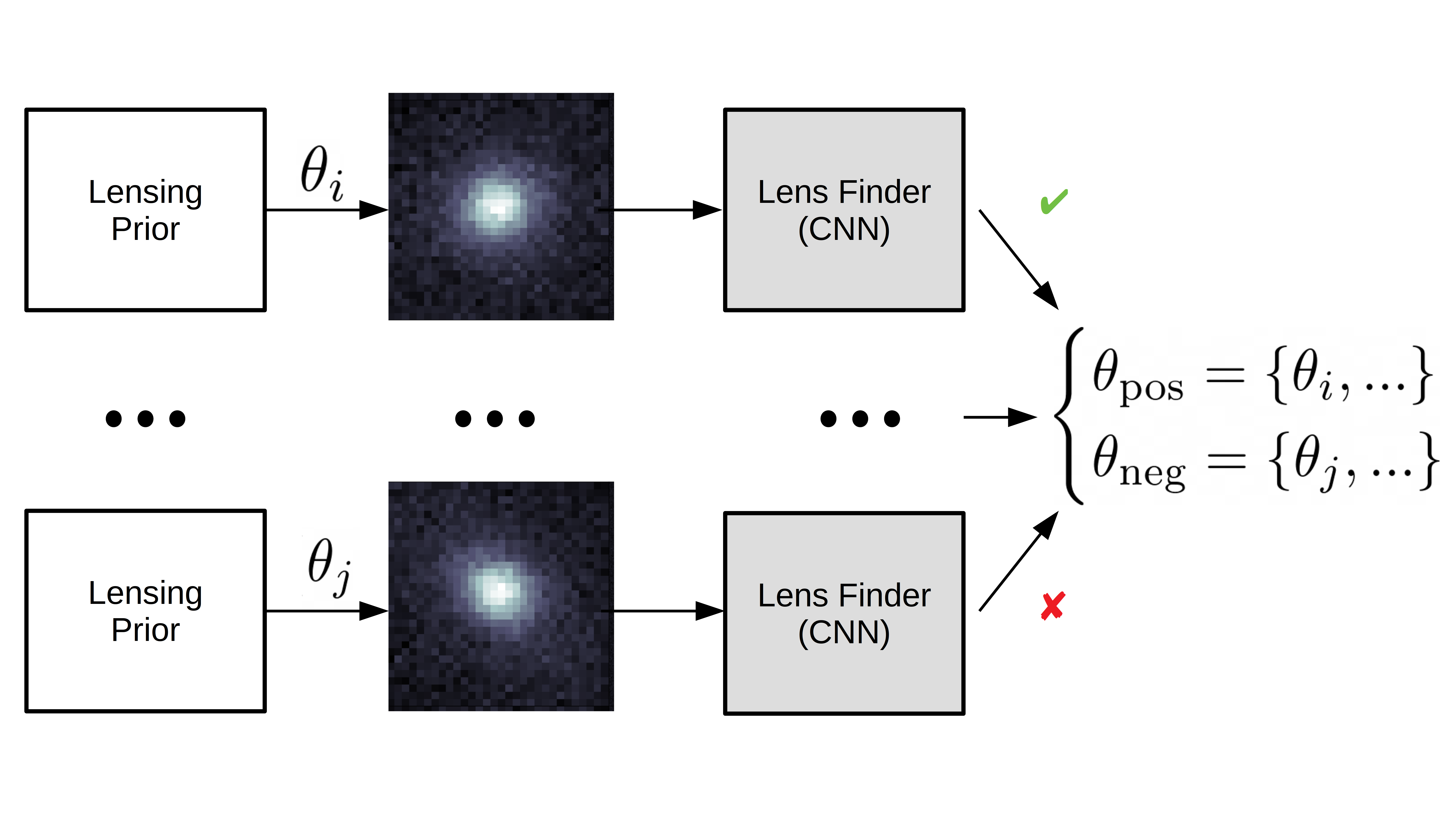}
\centering
\caption{Flow chart illustrating the generation of training data used for training the neural selection function. The CNN-based lens finder is applied to simulated low-resolution strong lensing images with model parameters $\theta$ sampled from a prior distribution. Two sets of model parameters are made based on whether the images were detected as strong lenses. The neural selection function is trained to classify between these two sets, where the output classification probability represents the probability of detection $p_{\mathrm{det}}(\theta)$.}
\label{flowchart1}
\end{figure}

To generate training data, we apply a CNN detector to a set of simulated low-resolution strong lensing images. This reflects the lens search procedure in LSST-like survey data. Lenses with parameters $\theta$ which are detected by the CNN are classified as detectable, and the missed lenses are classified as non-detectable. An illustration of the training data generation is shown in Figure \ref{flowchart1}.

The neural network that predicts the selection function $p_{\mathrm{det}}(\theta)$ is composed of an input layer and a hidden layer each with 256 neurons, in addition to a final layer outputting a single value. We use the Exponential Linear Unit activation function \citep[ELU]{2015arXiv151107289C} for the first two layers, and the sigmoid activation for the output of the network. The final layer predicts the probability of $\theta$ being classified as detectable, which gives us a prediction for the probability of detection $p_{\mathrm{det}}(\theta)$. The weights of the neural network are updated using the Adam optimizer with a constant learning rate of $10^{-4}$. Our training data consists of two equal sets of 10,000 detectable and non-detectable $\theta$ parameters. The network is trained for 10,000 epochs with a batch size of 1000.

\section{Results}
\label{results}

We compute the posterior $p(\lambda|\{x_i\})$ using Equation \ref{pop_posterior_final} from a set of 1000 CNN-detected strong lensing observations. As detailed in section \ref{methods_selec}, we first train a neural network to predict the CNN detector selection function $p_{\mathrm{det}}$. Then, we use the neural network's prediction of $p_{\mathrm{det}}$ in Equation \ref{alpha_eq} to compute the normalization factor $\alpha(\lambda)$ over a grid of $\lambda$ values. Separately, we train a BNN to infer the lensing posterior $p(\theta|x_i)$ for the individual strong lensing observations. In the final step, we use our prediction of $\alpha(\lambda)$ and $p(\theta|x_i)$ to compute the posterior of population hyperparameters $p(\lambda|\{x_i\})$ from Equation \ref{pop_posterior_final}. To approximate the expectation values from Equation \ref{pop_posterior_final}, we use 100,000 samples from the posterior distribution $p(\theta|x_i)$ for each observation $x_i$. We compute $\alpha(\lambda)$ by approximating the integral from Equation \ref{alpha_eq} using 10,000 samples from the population distribution $p(\theta | \lambda)$. The computation of $\alpha(\lambda)$ is repeated 1000 times using different samples from $p(\theta | \lambda)$ and averaged over in order to reduce noise in its prediction. This helps mitigate instabilities that may occur when dividing by the normalization factor $\alpha(\lambda)$ in Equation \ref{pop_posterior_final}. A flow chart illustrating the problem setup for inference is shown in Figure \ref{flowchart2}.

\begin{figure}
\includegraphics[width=8cm]{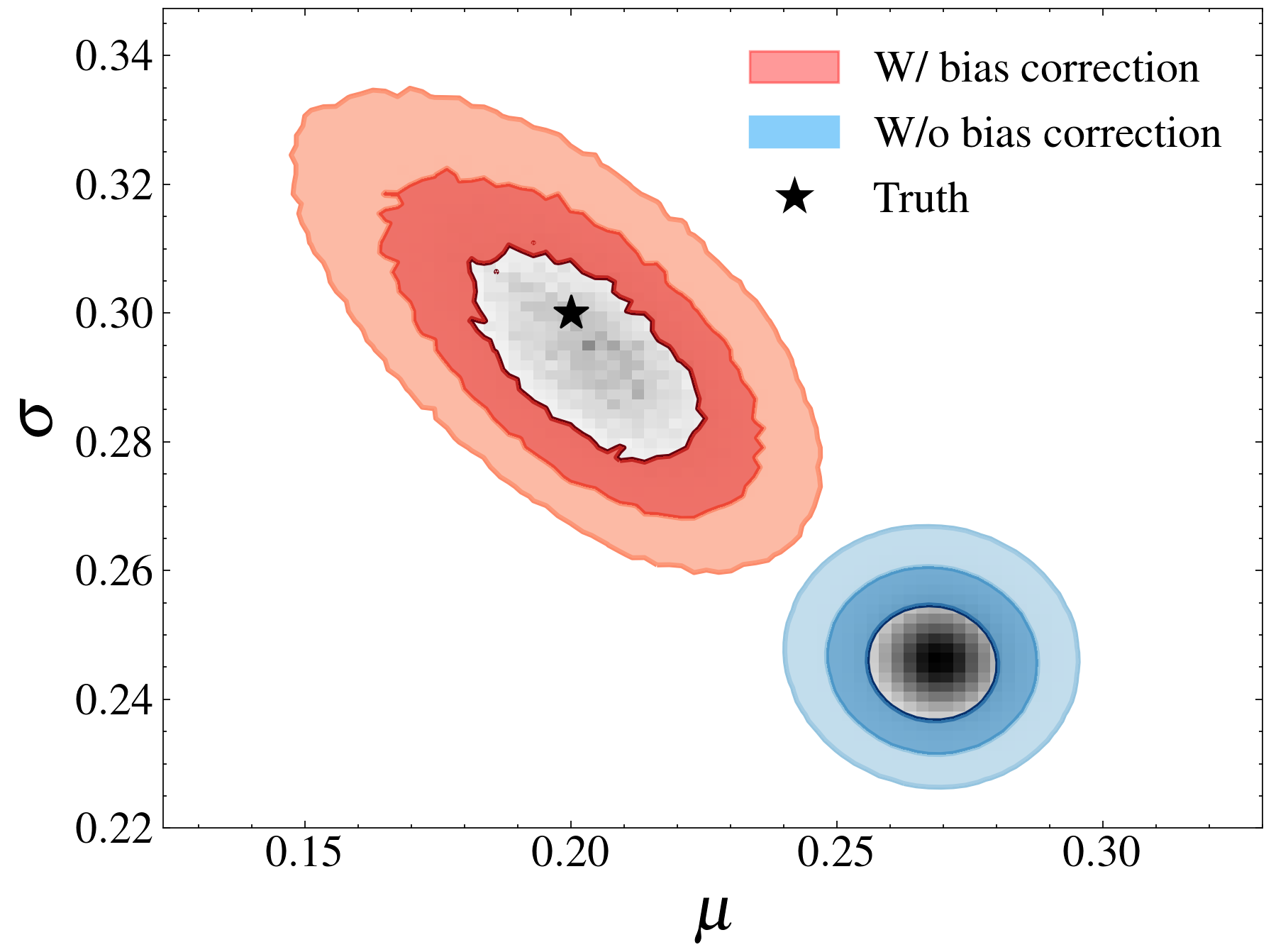}
\centering
\caption{Predicted posterior $p(\lambda|\{x_i\})$ given a set of 1000 strong lensing observations detected using the CNN strong lensing detector with the selection bias correction included within the hierarchical model (red contours) and without (blue contours). The $68\%$, $95\%$ and $99.7\%$ probability regions are shown for both posterior distributions. This result demonstrates that ignoring the effects of the selection bias caused by the CNN detectors leads to a biased inference of the population hyperparameters $\lambda$.}
\label{posterior_plot}
\end{figure}

For our problem, the hyperparameters $\lambda$ that we wish to infer are the mean $\mu$ and standard deviation $\sigma$ of a log-normal distribution describing the population distribution of strong lensing Einstein radii $R_E$. As an initial test, we apply a mask on the strong lensing light for lensing systems with Einstein radius smaller than $0.8$ arcseconds, making it impossible for the CNN finder to detect the presence of strong lensing. This leads to a more important selection bias towards systems with larger Einstein radii, and for the initial test, allows to more intuitively visualize the impact of selection bias on the final result. In Figure \ref{posterior_plot}, we show results of the predicted posterior $p(\lambda|\{x_i\})$ for two different scenarios: one in which the hierarchical model includes the correction due to selection bias and the other where it is ignored (i.e. whether $\alpha(\lambda)$ is included in Equation \ref{pop_posterior_final}). 

\begin{figure}
\includegraphics[width=8cm]{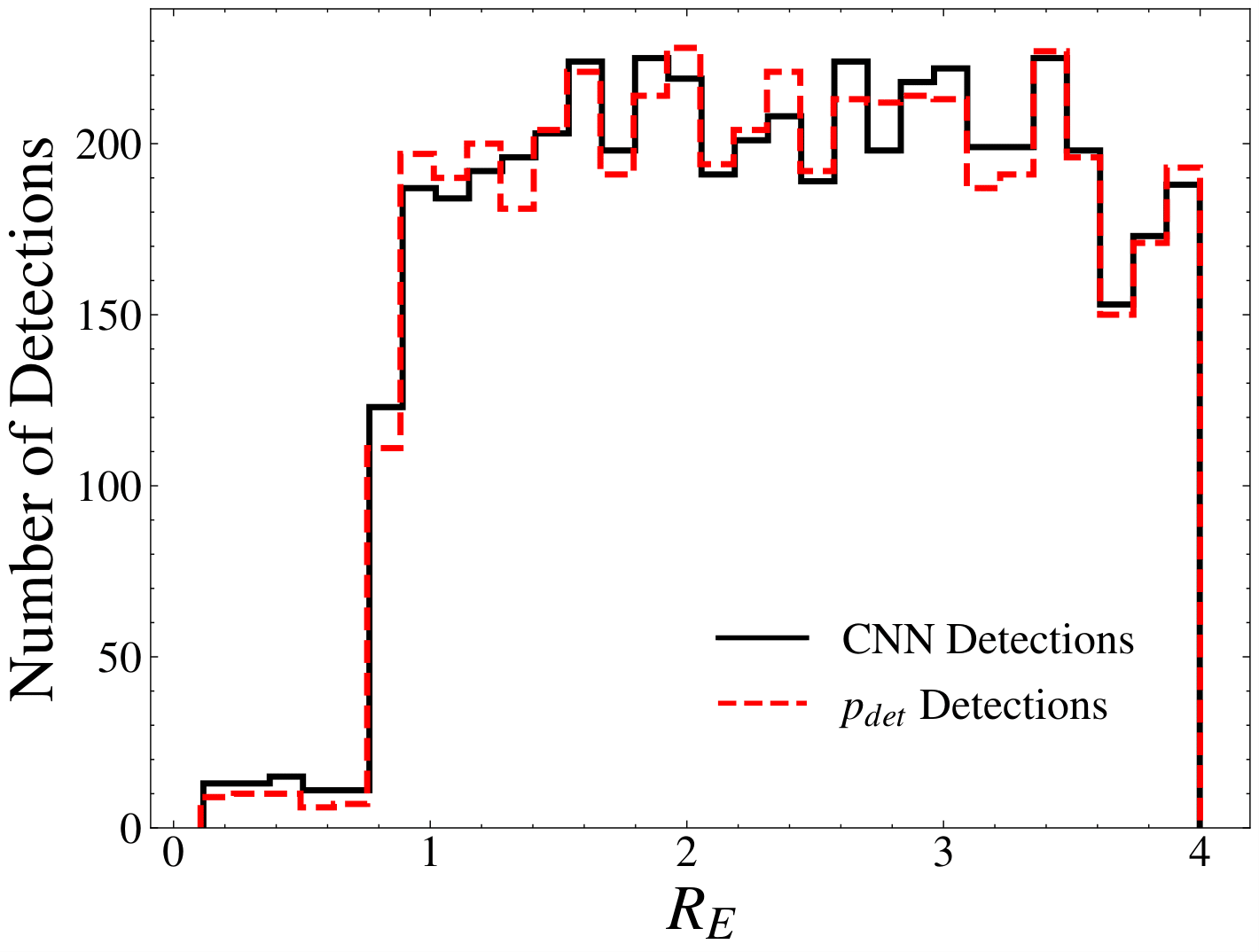}
\centering
\caption{The predicted number of strong lenses detected from a set of 10,000 simulated strong lenses by the CNN detector (black line) and by our modeled selection function $p_{\mathrm{det}}$ (dotted red line) as a function of the Einstein radius $R_E$. The neural selection function $p_{\mathrm{det}}$ is capable of accurately portraying the selection mechanism of CNN-based lens detectors.}
\label{rein_plot}
\end{figure}

We also perform a test to verify if our modeled selection function $p_{\mathrm{det}}$ accurately portrays the performance of the CNN strong lensing detector. To do so, we start by generating a new set of 10,000 strong lensing observations based on a wide range of model parameters $\theta$, including the Einstein radius $R_E$. Then, in the first case, we evaluate the detection probability $p_{\mathrm{det}}(\theta)$ for these $\theta$ samples, and keep the samples considered as detected if they satisfy $\rho < p_{\mathrm{det}}(\theta)$, where $\rho$ is a number randomly sampled between zero and one for each $\theta$. We denote this set of detected $\theta$ samples as $\theta_{p_{\mathrm{det}}}$. In the second case, we apply the CNN detector directly on the strong lensing images. For the observations that passed as a detection, we keep the samples of $\theta$ that were used to generate them. We denote this set of samples as $\theta_{\text{CNN}}$. In Figure \ref{rein_plot}, we plot the distribution of the Einstein radius parameter $R_E$ from both sets of $\theta_{p_{\mathrm{det}}}$ and $\theta_{\text{CNN}}$. This test allows us to verify if our modeled selection function $p_{\mathrm{det}}(\theta)$ accurately reflects the overall distribution of strong lenses that are discovered by the CNN detector.

\section{Discussion}

The framework presented in this paper allows for the hierarchical inference of population-level parameters of strong lensing systems, including the effects of selection functions. In the first stage of inference, the posteriors of the parameters of individual lensing systems are obtained. In the second stage, the distribution of the parameters of the general population of lenses is inferred using the posteriors of individual lens parameters. In this second stage of inference, it is also essential to include the effect of the selection function.

It is of course possible to forgo a hierarchical framework and infer population-level parameters in a one-stage inference setup. Simulation-based (or likelihood-free) inference methods that directly model target quantities such as $p(x|\lambda)$ have proven to be promising methods for inferring population-level parameters in cosmology \citep{2019ApJ...886...49B, 2020arXiv201007032C, 2021PhRvD.104h3531G}. 

However, Bayesian hierarchical modeling can be a flexible and powerful alternative, providing several advantages.
A distinct advantage of hierarchical modeling is the fact that the inference of the population-level parameters is done independently from the inference of individual systems. In this framework, the posterior of the lens parameters given the data $p(\theta|x_i)$ (the noise term) can be calculated once with expert (lensing) knowledge, either using machine learning based methods \citep[e.g.,][]{2017Natur.548..555H,2021ApJ...909..187W, 2021arXiv211205278L} or likelihood-based methods \citep[e.g.,][]{2022arXiv220207663G}. The inference of population-level parameters could then be carried out and explored without redoing any analysis of the observed data. The term $p(\theta|\lambda)$, often called the ``theory term'', describes the theoretical prediction of lens parameters under a proposed hyperparameter. This allows one to explore any theoretical model in a modular way using the same set of posteriors, $p(\theta|x_i)$, calculated earlier. 

In this work, we modeled the selection function $p_{\mathrm{det}}(\theta)$ with neural networks, facilitating accurate calculation of the posterior of the hyperparameters without the need for repeated Monte Carlo simulations. Once learned, the same neural selection function could be used to recalculate the denominator of Equation \ref{biased_likelihood} for any new theoretical model under consideration.

The approach proposed here uses simulation-based inference and machine learning models to learn surrogate models for the various modules that form a hierarchical inference model. This allows taking advantage of the power and the versatility of neural networks while retaining the benefits and the interpretability of Bayesian hierarchical frameworks, where uncertainties are correctly propagated from one level to the next.

\section*{Acknowledgements}
This research was supported by a generous grant from the Schmidt Futures Foundation. The work was also enabled in part by computational resources provided by Calcul Quebec, Compute Canada and the Digital Research Alliance of Canada. Y.H. and L.P. acknowledge support from the National Sciences and Engineering Council of Canada grant RGPIN-2020-05102, the Fonds de recherche du Québec grant 2022-NC-301305, and the Canada Research Chairs Program. We acknowledge support from the Natural Sciences and Engineering Research Council of Canada (NSERC), [funding reference number 411299547]. We thank Adam Coogan and Alexandre Adam for useful discussions in regards to simulation-based inference methods.

\bibliography{bibliography}
\bibliographystyle{icml2022}

\newpage
\appendix
\onecolumn
\section{Appendix}
\label{appendix}
 \begin{figure}[h]
 \includegraphics[width=12cm]{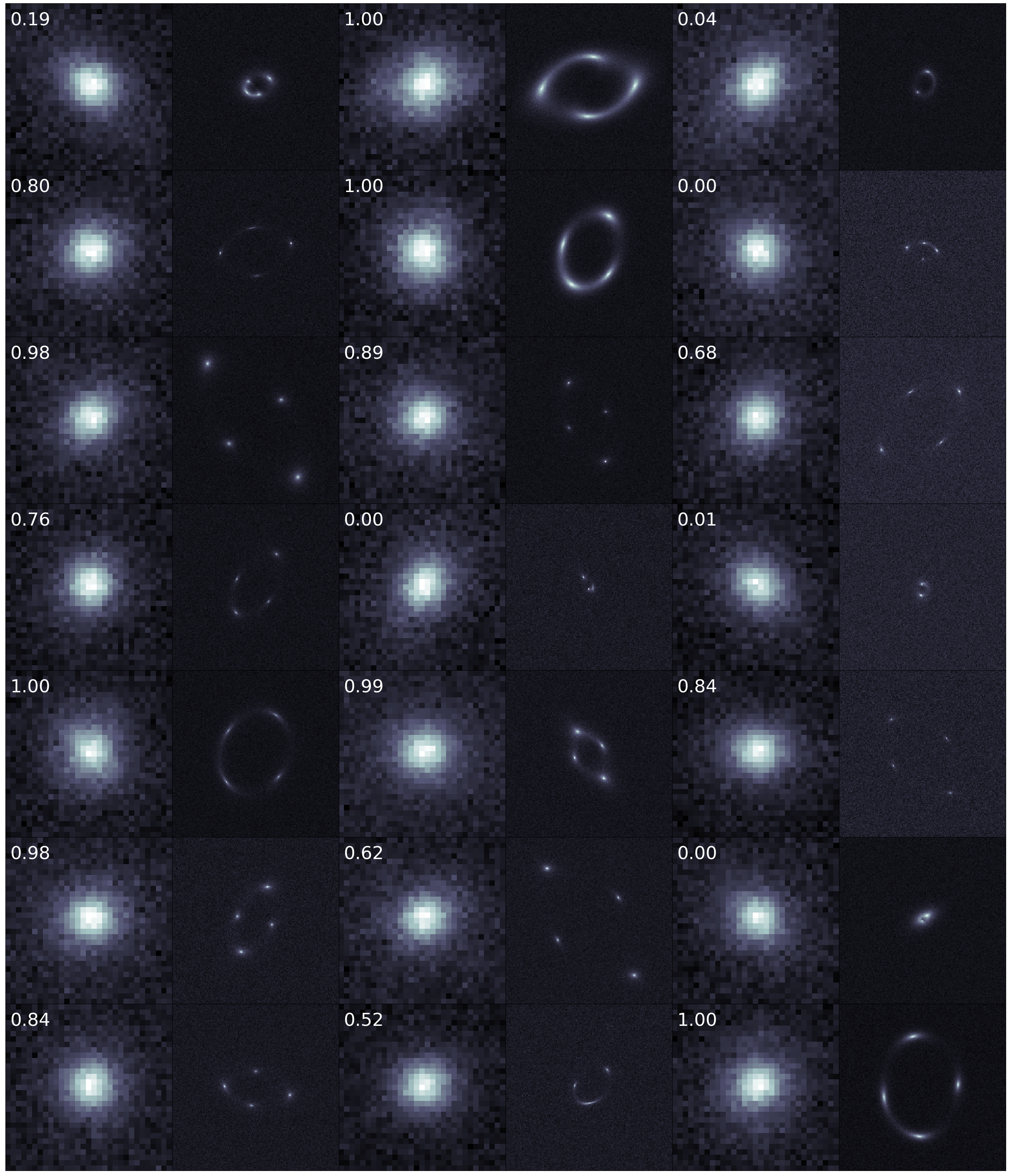}
 \centering
 \caption{Examples of generated strong gravitational lensing simulations. The set of simulations used consists of low-resolution wide-area sky surveys (left) and the corresponding follow-up high-resolution image (right). The probability of detection $p_{\mathrm{det}}$ predicted by our neural network model is shown in the top corner of each low-resolution image. The strong lensing simulations are composed of a Singular Isothermal Ellipsoid \citep[SIE]{1994A&A...284..285K} with added external shear representing the main lens deflector and a background source following a S\'ersic brightness profile \citep{1963BAAA....6...41S}. We also add stellar light following a S\'ersic profile to the main lens deflector galaxy in the low-resolution images. This is not done for the high-resolution follow-up images as we assume that the stellar light from the main deflector galaxy is removed in a data preprocessing stage. As a whole, our simulations can be fully described by a 13-dimensional parameter vector $\theta$.}
 \end{figure}

\begin{figure}
\includegraphics[width=8cm]{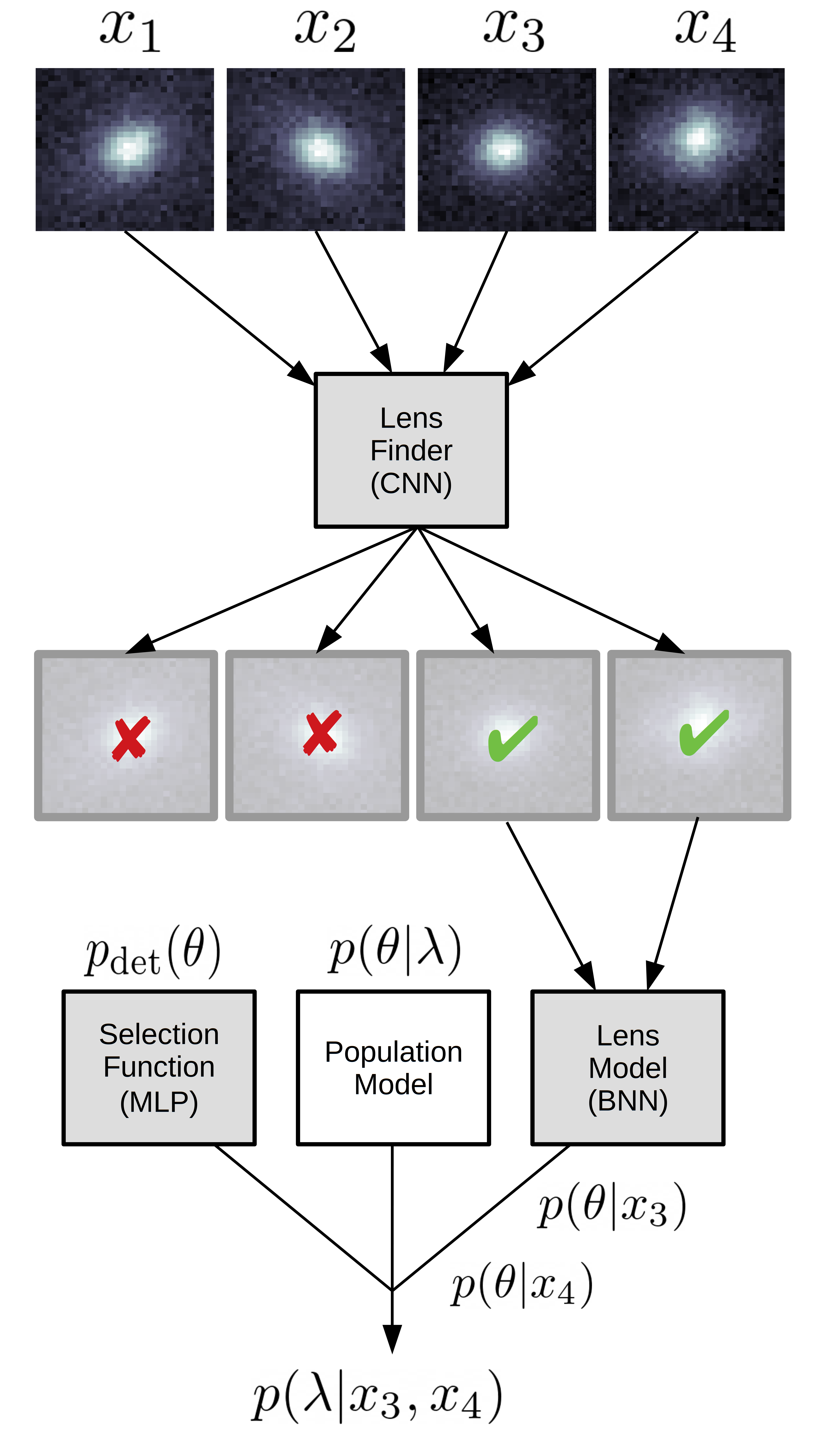}
\centering
\caption{Flow chart illustrating the general setup for the inference of the population hyperparameter posterior $p(\lambda|\{x_i\})$ for an example with four generated strong lensing observations $\{x_1, x_2, x_3, x_4\}$. The grey colored boxes are the neural networks used to perform the hierarchical inference task. The lens finder (CNN-based detector) is applied to the observations and classifies whether they contain strong lensing. For the images detected as strong lenses, the posterior of lensing model parameters $p(\theta|x_i)$ is obtained by applying a Bayesian Neural Network (BNN) on the followup high-resolution images. The population distribution $p(\theta|\lambda)$, neural selection function $p_{\mathrm{det}}(\theta)$ and predicted lensing posteriors (in this example, $p(\theta|x_3)$ and $p(\theta|x_4)$) are used to compute the posterior of population-level parameters $p(\lambda|x_3, x_4)$.}
\label{flowchart2}
\end{figure}

\end{document}